# The time is ripe to reverse engineer an entire nervous system: simulating behavior from neural interactions

Gal Haspel (NJIT), Ben Baker (Colby College), Isabel Beets (KU Leuven), Edward S Boyden (MIT), Jeffrey Brown (MIT), George Church (Harvard University), Netta Cohen (University of Leeds), Daniel Colon-Ramos (Yale University), Eva Dyer (Georgia Institute of Technology), Christopher Fang-Yen (Ohio State University), Steven Flavell (MIT), Miriam B Goodman (Stanford University), Anne C Hart (Brown University), Eduardo J Izquierdo (Rose-Hulman Institute of Technology), Konstantinos Kagias (MIT), Shawn Lockery (University of Oregon), Yangning Lu (MIT), Adam Marblestone (Convergent Research), Jordan Matelsky (University of Pennsylvania), Brett Mensh (Optimize Science), Talmo D Pereira (Salk Institute), Hanspeter Pfister (Harvard University), Kanaka Rajan (Harvard Medical School), Horacio G Rotstein (NJIT), Monika Scholz (Max Planck Institute for Neurobiology of Behavior), Joshua W. Shaevitz (Princeton University), Eli Shlizerman (University of Washington), Quilee Simeon (MIT), Michael A Skuhersky (MIT), Vineet Tiruvadi (Hume AI), Vivek Venkatachalam (Northeastern University), Donglai Wei (Boston College), Brock Wester (Johns Hopkins APL), Guangyu Robert Yang (MIT), Eviatar Yemini (UMass), Manuel Zimmer (University of Vienna), Konrad P Kording (University of Pennsylvania)

**Abstract:**

Just like electrical engineers understand how microprocessors execute programs in terms of how transistor currents are affected by their inputs, neuroscientists want to understand behavior production in terms of how neuronal outputs are affected by their inputs and internal states. This dependency of neuronal outputs on inputs can be described by a state-dependent input-output (IO)-function. However, to reliably identify these IO-functions, we need to perturb each input and combinations of inputs while observing all the outputs. Here, we argue that such completeness is possible in *C. elegans*; a complete description that goes all the way from the activity of every neuron to predict behavior. The established and growing toolkit of optophysiology can non-invasively capture and control every neuron's activity and scale to countless experiments. The information from many such experiments can be pooled while capturing the inter-individual variability because neuronal identity and function are largely conserved across individuals. Just like electrical engineers use transistor IO-functions to simulate program execution, we argue that neuronal IO-functions could be used to simulate the impressive breadth of brain states and behaviors of *C. elegans.*

**Introduction**

When electrical engineers design microprocessors, they use simulations that start with knowledge of how the components, primarily transistors, work and simulate them over time. These simulations are thus key to designing microprocessors and help engineers understand the processors in depth. In fact, older microprocessors sometimes contain elements designed to make it hard for others to reverse engineer the microprocessor and successfully simulate it (like the 6502 (Jonas and Kording 2017)). For such a simulation, one needs a list of all components (netlist) and the input-output (IO)-functions of all the elements. The simulation is then the key tool for understanding information processing in the microprocessor and asserting that it is correctly engineered.

When philosophers talk about understanding and explaining, they usually distinguish many kinds thereof. According to Niko Tinbergen (Tinbergen 1963), there are four categories of explanations, divided along the axes of proximate to ultimate and static to dynamic. One goal is to understand the evolutionary trajectory (phylogeny) that made the nervous system the way it is over long timescales. Another goal is to understand how development (ontogeny) gave rise to the individual animal as it is now. A third goal is to understand the function (adaptation) of the nervous system. And lastly, the goal is to reveal mechanisms (causation). This goal is usually taken as describing the function of something in terms of causal interactions between the involved components. Just like electrical engineers explain the function of the microprocessor in terms of interactions between transistors, chemists understand how the interactions between atoms give rise to chemical bonds and reactions, and physicists understand how interactions between subatomic particles and forces give rise to matter. Similarly, neuroscientists want to understand how the interactions among sensors, neurons, muscles, and other building blocks give rise to behavior. This reductionist focus on understanding emergent behavior in terms of the mechanisms mediated by components is a key goal of much of science.

When neuroscientists aim to understand brains, following Tinbergen's four questions, they thus use many different levels and definitions of "understand." For some, understanding means the ability to fix diseases. For others, understanding means knowledge of involved representations. And yet, for some, understanding means the ability to simulate. In analogy to the electrical engineering case, the ability of faithful simulation would then also come with solutions for fixing problems, predicting the nature of representations, and a range of approaches to deepen our understanding of the brain.

Establishing a mechanism generally requires the establishment of a causal link. For example, we want to know that a neuron causes another neuron to fire more often via an excitatory synapse between the neurons. Establishing causality is often difficult, and a well-developed branch of statistics focuses on such problems. This field has figured out precise definitions of the situations when causality can be inferred (Pearl 2009), and the basic upshot is that establishing causality generally requires perturbations (Woodward 2005; Pearl 2009). Establishing the causal influence of one neuron on another will thus generally require us to stimulate the former and record from the latter.

Just like we need to simulate all transistors to simulate an entire microprocessor, we need to simulate all neurons to simulate an entire nervous system. We need a completeness of simulation to understand how the components mechanistically integrate to produce behavior. We need a completeness of recording because we need the IO-function of each neuron and completeness of perturbations to establish causality (Woodward 2005; Pearl 2009) and reach statistical power (see below). Hence, we need perturbation and recording completeness to enable complete simulations to enable a complete mechanistic understanding.

Here, we argue that this completeness can now be obtained in the nematode *C. elegans*. The animal behavior can be recorded and fully described in low dimensions (Stephens et al. 2008) or a sequence of a finite number of poses (Costa et al. 2024); all neurons and muscle cells can be recorded (Nguyen et al. 2016; Uzel, Kato, and Zimmer 2022); and an arbitrary subset can be stimulated or inactivated (M. Liu et al. 2022; Pokala et al. 2014; Schmitt et al. 2012). Therefore, it should be possible to identify the IO-function of every neuron (just like the corresponding function for transistors). Given that we already know and can identify all the neurons, glia, and muscle cells (just like the netlist of the microprocessor), knowing all cellular IO-functions should be enough to enable simulating the network IO-function and the animal behavior. We also spell out what this would take and why its completion would be important for neuroscience.

**Why should we reverse engineer a nervous system?**
"What I cannot create, I do not understand," one of Richard Feynman's famous dictums, highlights the need to build systems to drive their understanding (also see (Dretske 1994)). Reverse engineering nervous systems by characterizing the state-dependent IO-functions of neurons and muscle cells may enable us to recreate and understand neural systems driving complex behaviors.

A central goal of systems and computational neuroscience is to model how nervous systems convert stimuli, spontaneous activity, and internal states into coordinated muscle contractions, which are behaviors (also see (Harel 2003; Krakauer et al. 2017)). Indeed, the National Institute of Health and other funders started the BRAIN Initiative (Insel, Landis, and Collins 2013) with a multi-billion dollar investment to develop new large-scale neurotechnologies. Major initiatives by other funders include the Human Brain Project (Markram 2012), MICrONS (funded by the Intelligence Advanced Research Projects Activity), and the Simons Global Brain Collaborations. The investment of significant resources into reverse engineering the brain reflects the value of this endeavor.

Let us be clear about what we mean by successful reverse engineering. Independent of variations in cellular biophysics, we consider each neuron and muscle cell to be computing its output as a function of the activity of its input cells and spontaneous activity. As such, it is characterized by a spatiotemporal function mapping input traces into an output trace (i.e., an IO-function), which represents both synaptic and nonsynaptic effects (e.g., neuropeptidergic signaling or spontaneous activity). In this reductionistic framing, reverse engineering consists of figuring out the input-output mapping for all neurons and muscle cells as well as the inputs from the world (i.e. system identification), then reassembling the collection of IO-functions into a robust, simulatable model that can exhibit key behaviors when connected to the simulated body (Fig. 1). Moreover, behavioral variability will be recapitulated by drawing parameters for multiple simulations from the empirical variability of IO-functions. To be deemed successful, this working model should recapitulate behavior and behavioral variability under a range of conditions, stimuli, and perturbations. With a working model of a nervous system and its interactions with the body and world, we could test hypothesized neuroscientific models and principles in silico inexpensively and rapidly (Doerig et al. 2023).

**What will we learn by reverse engineering a nervous system?**
If we could decompose a simulation of the nervous system from neuronal IO-functions, we could predict downstream behavior in response to any sensory signals (past and present) under any experimental manipulations and internal states. We could then predict the full behavioral repertoire and what every neuron does as a function of brain state and stimulation. This ability to build a simulation is one of the definitions used for understanding in the systems neuroscience community (Kording et al. 2020) and is a natural product of reverse engineering (Csete and Doyle 2002).

But, falsifying or validating the in silico models is only the first step. Ultimately, reverse engineering a nervous system aims to build an explanatory model of the dynamics of a complete nervous system that captures adaptation and plasticity at cellular, circuit, whole-brain, and behavior levels.  A validated model that is consistent with all our data under some condition would provide a platform for running in silico experiments to test – precisely, specifically, and efficiently – the role of different model constituents (a neuron, a compartment, a connection, even a neuromodulator or ion channel) in neural and circuit dynamics. Such in silico experiments would then allow us to build understanding: to interpret the dynamics in terms of computational concepts, from decision-making, memory, and sensory integration to attention and coordination, and indeed to understand fundamental principles of circuit structure and function. This link between neuronal dynamics (both in the form of input-output relations and whole-brain states) and interpretable functions is arguably the holy grail of systems and computational neuroscience (Csete and Doyle 2002).

Being able to simulate the full nervous system of *C. elegans* may allow us to design new biologically-inspired systems to solve technical problems. The newfound knowledge could be used to inspire and develop therapeutic hypotheses for genetic disorders as well as for potential treatments tested through simulation. Such a model may catalyze the design of new information processing systems and intelligent signal processing systems that solve well-defined goals. The results may also inspire a new generation of artificial intelligence systems that are orders of magnitude more efficient than current ones (Abbott et al. 2020). In contrast to artificial intelligence systems, the nervous system of *C. elegans* is compact and incredibly energy efficient, and yet able to allow the species to thrive (Barrière and Félix 2005). As such, we may hope to find design principles that we can generalize to AI systems. Of particular interest here are the basic building blocks, figuring out which computational elements are used in *C. elegans,* e.g., in terms of nonlinearities and timescales, promises to inform the design of new low-energy AI systems as well as to make them more resilient in an adversarial world (Agarwal et al. 2017).

By reverse engineering any entire nervous system, we would gain important insights about the scientific process of biological reverse engineering that could generalize to larger systems, discovering which information matters and which shortcuts are possible. We now know synaptic patterns of connectivity (White et al. 1986; Cook et al. 2019; Witvliet et al. 2021; Brittin et al. 2021) poorly predict interactive patterns of neural and muscle activity (Bentley et al. 2016; Yemini et al. 2021; Susoy et al. 2021; Uzel, Kato, and Zimmer 2022; Beets et al. 2022; Ripoll-Sánchez et al. 2022; Dag et al. 2023; Atanas et al. 2023; Lu et al. 2022), so what are we missing to understand dynamic operations in these circuits? We know that, as in other animals, in *C. elegans,* individual neurons can compartmentalize signals and thus perform multiplex computations (Hendricks et al. 2012). So what resolution do we need to distinguish these compartmentalized signals? Do millisecond timescales matter, or is it enough to have lower frequency signals? Can we solve these problems with optical imaging only? Can the information describing the dynamics of nervous systems be compressed into a small number of principles? Is it enough to reverse-engineer using data on only some parts from each individual? We know that the biophysical properties of different cell types matter (Dag et al. 2023), but to what extent must these be understood and modeled? We now know that even in *C. elegans* with its invariant cell lineage, the wiring of neural circuits varies across individuals (Brittin et al. 2021; Witvliet et al. 2021). How can individual variability be taken into account and might data from many animals produce a meaningful general model? Does covariability matter? Does degeneracy matter?  We cannot currently answer these questions because we have not performed the required experiments and nervous system modeling. Pioneering reverse engineering in *C. elegans* can point out favorable approaches when we try to reverse engineer the more complex nervous systems of rodents and other animals. Demonstrating reverse engineering of an entire nervous system would clarify what kind of methods, data, and theory we may need for future, increasingly ambitious endeavors.

In reverse engineering a nervous system, we may also learn about failure modes. Can we easily be misled and believe we understand how a nervous system works from partial recording? How probable is it that the models we fit get the correlations right and the causality wrong (Tremblay et al. 2022)? How much data of what kind is too little to reverse engineer systems? Answers to these questions could be obtained for *C. elegans* and guide research in all areas where the goal is to understand nervous systems. Reverse engineering a system may also lead us to discover misleading "principles" in past neuroscience research.

By reverse engineering a nervous system, we will galvanize and motivate the enrichment and expansion of technologies that are critical for research across neuroscience. Neuroscience can scale: We can record from thousands of neurons, whereas we used to record from a handful (Stevenson and Kording 2011). We would optically image a million neurons where we previously imaged hundreds (Abdelfattah et al. 2022). Similar progress happened in molecular techniques where, for example, driving the human genome project ultimately made sequencing a cheap tool used routinely for countless objectives; analogously reverse engineering nervous systems should produce broadly useful techniques. Reverse engineering requires the development of hardware automation, data handling, and sharing, as well as algorithms and software to extract and analyze neural activity from dense whole-brain recordings, yielding a suite of interlocking technologies that can be scaled up and potentially generalized. Integral to this vision would be an informatics infrastructure enabling an unprecedented and integrated data distribution of connectomics, neural activity, behavior, and computational models.

The existence and fact-checking of all these resources would jump-start a drive toward scaling entire nervous systems' reverse engineering and simulation.

**Why have we not yet reverse-engineered a nervous system?**
Experimental limitations and theoretical realities mean that, despite earlier efforts (Sarma et al. 2018), we have yet to simulate the entire nervous system of *C. elegans*. We need many parameters to describe *C. elegans* and, astronomically many to describe mammals. Our experiments so far have been too limited in breadth, but also, focusing on correlations, have not measured the causal parameters needed for simulations. Indeed, it remains an open and hotly debated question whether and how theories can supply the missing parameters.

One reason why these challenges appear insurmountable is that we focus on reverse engineering rodents, with most of the circuit work in mice. However, we can only observe a tiny part of their nervous system; neither can we record all inputs to a single neuron. There is no doubt that the relevant models for mice are exceptionally complicated (Abbott et al. 2020), both in terms of neuronal and neural processing and in terms of learning during experimental procedures. Because neuroscientific research takes place mostly on mammals, it is not known how good modern techniques may be at reverse engineering these circuits (Jonas and Kording 2017). After all, individual neurons are often weakly correlated with behavior and the enormous numbers of neurons in mammals (about 86 billion neurons in human and 70 million in mice (Herculano-Houzel, Mota, and Lent 2006)), preclude a population-level view with the single-neuron resolution necessary to understand their in-depth circuit dynamics. Moreover, every cortical neuron in the rodent receives input from thousands of neurons on the same order of magnitude as all the neural connections in the nematode nervous system. The *Drosophila* nervous system, with about 130,000 neurons, is a productive model for systems neuroscience, revealing important features of how circuits control complex behaviors, like flexible navigation or learning and memory (Fisher 2022; Modi, Shuai, and Turner 2020). However, even with extensive emerging connectome data (Scheffer et al. 2020; Dorkenwald et al. 2023) and transgenic tools, it is not yet feasible to record all neurons with knowledge of cellular identity or stimulate every neuronal cell type in succession during brain-wide recordings. Such datasets may be essential to build accurate simulations of a nervous system. In summary, in the field's quest to understand complex nervous systems we have started with some of the most complicated ones. Not only do we not know how to simulate such nervous systems, we do not know what we do not know, which model components we need, and which data we need to constrain those components on the path to simulate them.

We have had much of the *C. elegans* connectome since 1986 (White et al. 1986); further annotated (Varshney et al. 2011; Chen, Hall, and Chklovskii 2006; Cook et al. 2019); and new datasets were added (Brittin et al. 2021; Witvliet et al. 2021; Mulcahy et al. 2022). But connectivity alone is insufficient because without knowing the strength and temporal properties of all connections and neuronal properties (Kopell et al. 2014), "... it would not be possible to simply go from the wiring diagram to the dynamics of even two neurons" (Bargmann and Marder 2013). Despite impressive progress (Einevoll et al. 2019; Eliasmith and Trujillo 2014), we cannot yet simulate any entire nervous system.

No discussion of reverse engineering *C. elegans* can be complete without careful consideration of the OpenWorm project (Szigeti et al. 2014)(https://openworm.org/). The OpenWorm project impressively used relatively solid knowledge about the physical environment of the nematode, the physics of swimming, and muscle properties (Boyle and Cohen 2008; Boyle, Berri, and Cohen 2012). They used the very partial relevant published information (White et al. 1986; Varshney et al. 2011) as well as a two-dimensional atlas (Hall and Altun 2008) of nematode nuclei. Impressively, OpenWorm even began compiling a list of ion channel inventories for neurons and models of their biophysical properties (Sarma et al. 2018). However, estimating neuronal interactions from behavior, or even from ongoing neuronal activity, is essentially impossible, as many different neuronal properties can produce the same behavior (Prinz, Bucher, and Marder 2004). Moreover, in the absence of standardized data formats and norms for data sharing, the transfer of neural data into published papers is equally an irreversible process. By starting the simulation project, the OpenWorm project took a massive step in the right direction. But missing are experiments explicitly aimed at producing the kind of data that would allow a faithful simulation. Data was the limiting factor all along.

Rapidly developing advances in data-driven approaches, including artificial intelligence, combined with growing computational power, provide much-needed tools for learning multidimensional, potentially non-linear mappings for IO-relationships. Concurrently, key experimental technologies, including microscopy, optophysiology, experimental automation, and data management, have reached the required capabilities. The tractability of *C. elegans* and the advent or maturation of these tools for reverse engineering converge into a unique opportunity.

**Why a small nervous system? Why *C. elegans*?**
*C. elegans* has far fewer neurons and neuron-neuron connections than other model animals making it a great starting point to measure the causal dynamics, captured by its IO-functions, driving stereotyped behavior. Moreover, all muscle cells can also be recorded in live behaving animals to compile a complete description of the motor output that generates behavior. It has established optical techniques for recording and stimulating at scale (Nguyen et al. 2016; M. Liu et al. 2022; Bergs et al. 2023; Randi et al. 2022b). Importantly, all *C. elegans* neurons, glia, and muscle cells are identifiable, allowing us to combine information across animals. Until recently, variations among nematodes were thought to be relatively unimportant, and physiological parameters were considered relatively conserved (Randi et al. 2022b), supporting the pooling of data across individuals. More recent evidence of variability in the neuronal wiring (Brittin et al. 2021), neuronal encoding (Atanas et al. 2023), and behavior (Stern, Kirst, and Bargmann 2017) across individuals suggests a role for individuality and compels addressing variabilities when conducting simulations. Put together, *C. elegans* offers a unique system to address these questions: by analyzing many individuals, we can determine the extent of variation at the cellular, circuit, and behavioral levels, and we can take these variations into account in building computational models. Simulation might be facilitated by the electrical signals common in *C. elegans* requiring lower temporal resolution of both stimulation and recording (M. B. Goodman et al. 1998; Jiang et al. 2022; Q. Liu et al. 2018; Lockery and Goodman 2009). From the developmental lineage, through molecular pathways, to the first connectomes, *C. elegans'* important contributions to scientific knowledge mainly emerged from research approaches that have aimed for completeness. For example, sequencing the *C. elegans* genome also significantly advanced biology and laid the groundwork for sequencing the human genome (C. elegans Sequencing Consortium 1998).

*C. elegans* is small enough to perturb and record calcium signals from each neuron. Continuous recording and stimulation on a single setup would reach about 50,000 specimens a year and can be parallelized across setups (Fig. 1AB). Undoubtedly this is a large number, but it is only an order of magnitude larger than the number of animals whose behavior was recorded for recent publications (Dag et al. 2023). Experimental advantages allow a very large number of experiments, clearly enough for at least measuring the effect of all pairwise stimulations (~90k pairs). In *C. elegans* and only in *C. elegans*, this approach can sample all combinations of presynaptic neurons. In other words, within a single year, it is possible to densely sample the space of inputs to all the neurons and muscle cells.

**What will reverse engineering the *C. elegans* nervous system deliver?**
The result of reverse engineering a brain is a dynamical model from which one could generate testable predictions for all possible manipulations and experiments. In other words, we want a causally correct 'digital twin' of the biological system. A cohort (e.g., 1.000) of variable twins should react to any kind of stimulation in the same way as a cohort of real individuals. So what is the minimal aspect of what we mean by wanting to reverse engineer it? Let us go through the outcomes we want to have.

The model should be able to describe the nervous system and its motor output (i.e. the patterns of muscle contractions that constitute a behavior). This description will have to include the influence of each neuron on other neurons and ultimately on behavior to predict stimulation effects. Importantly, neurons interact through synapses with neurotransmitters, but also through gap junctions, neuromodulators, direct electrical coupling (Anastassiou et al. 2011), and non-synaptic and non-neuronal paths such as glia (Raiders et al. 2021; Mu et al. 2019), as well as muscle activations and body shape (Cohen and Denham 2019; Zhan et al. 2023). We thus need a model that captures interactions of these kinds. It should be able to simulate the entire trajectory of neural activities and thus predict activities and tuning curves at all times and contexts (Hallinen et al. 2021).

While our conceptualization is built on there being little structured natural variability in wiring, neural activity, and behavior among individuals, the measured variability in I-O functions should be included in the model and should replicate the variability in behavior. Importantly, by conducting enough experiments, we can measure both average properties and their overall distribution or, indeed, identify multiple solutions. The behavioral variability among hundreds of simulations drawn from the measured distribution should be similar to that among hundreds of animals.

The model should be able to produce all the basic behaviors the animal performs, including spontaneous, responsive, and coupled behaviors. The hermaphrodite *C. elegans* has an impressively rich behavioral repertoire (Hart 2006; Atanas et al. 2023), constructed of first-order building blocks including: 1) locomotion behaviors, including forward, backward, change of speed, steer, turn, and halt; 2) feeding behaviors, more specifically occurrence, rate, and coordination of pharyngeal pumping (used to collect and crush organic material such as bacteria); 3) defecation; 4) egg laying in hermaphrodites; 5) fixed action patterns for mating in males. These behaviors can all be studied effectively with minutes-long recordings, and hence the above list excludes behaviors with a longer timescale to currently doable (and scalable) experiments (e.g. learning). We envision a simulation that replicates these building blocks of all behaviors in a single model. Doing so requires a simulation of the whole body, the inputs from the world around it, and how its effectors, in turn, affect the world around them, at least at a level that allows interpretation of motor output to behavior. In our opinion, these are the minimal objectives for producing acceptable models of the nervous system. We expect that such a simulation should replicate the more complex behaviors that are made out of the first-order ones (Ghosh et al. 2016; Dekkers et al. 2021), as well as the operating system that chooses amongst the many possible first-order behaviors available at a given time, to confront the complexity of a dynamic and complex environment. These more complex behaviors include finding and attracting mates, fleeing predators, seeking and avoiding chemical cues or temperatures, collective behavior, or avoiding pathogens and parasites. In other words, a good simulation should cover all the ethologically relevant behaviors described for the nematode.

**Feasibility: What do we need to do to reverse engineer a nervous system?**
We need a model of how the nematode's sensors translate the state of the nematode's body and the environment into neural activity, how neurons interact, and how neural activities through muscle contraction (effectors) influence the nematode's body and the environment. To close the loop, we also need a model for the environment (world). So we need a model of the world, the sensors, and the effectors (Fig. 1C)**.** With a model of the body and environment in hand, we should be able to calculate the information going into the animal at all times and how the action of the effectors produces its future dynamics (Fig. 1D). To get there, we primarily need a sequence of experiments and modeling that require the components below. Because each need can be achieved by multiple technologies and approaches, we will describe the need and give a few approaches and a concrete solution.

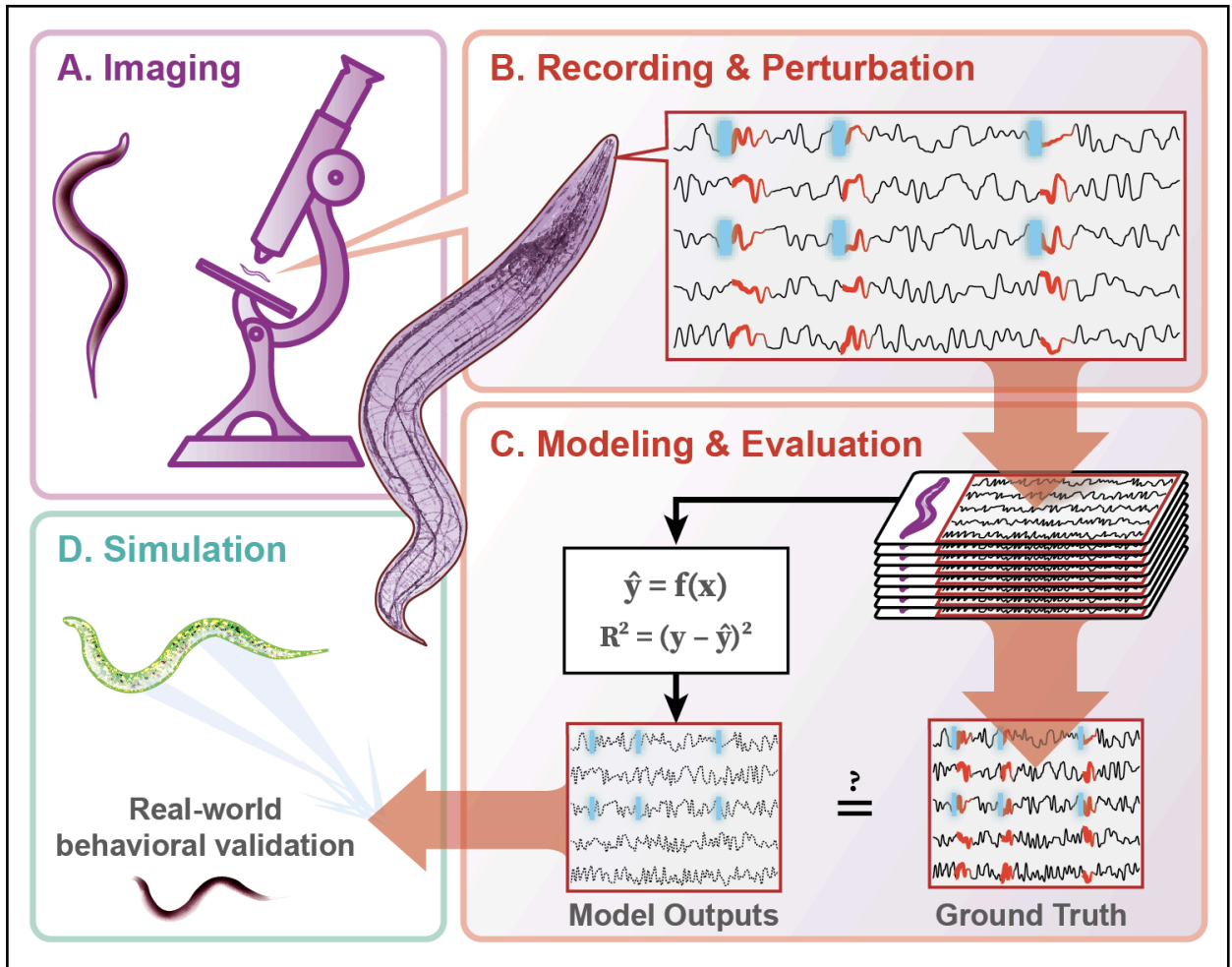

Figure 1: **An overview of the proposed approach, demonstrating how imaging, recording, perturbing, modeling, and simulation interplay.**

**(1) Automated staging**

*Need:* Any approach we can think of will require many animals, as we will need many thousands of hours of experiments to obtain the necessary statistical power to meaningfully simulate all neurons. We thus need a way of getting nematodes in and out of set-ups with high throughput. If we were to do that manually, we would require large amounts of human labor. Another complication would be to standardize a manual workflow, so automatic staging is a better option.

*Potential approaches:* Depending on the experimental design, animals could be immobilized, restricted, or freely moving. Each option introduces challenges in collecting and interpreting the data but all past experiments have used extensive human labor. Automation could include robotic manipulation (Li et al. 2022), microfluidics approaches that either move nematodes in an aqueous solution or small bubbles of oil (San-Miguel and Lu 2013), and machine-vision-based focusing and positioning (Li et al. 2022). Neuronal calcium imaging data recorded in freely moving animals (Nguyen et al. 2016; Venkatachalam et al. 2016; Susoy et al. 2021; Atanas et al. 2023), by their nature, do not reveal causality. The publicly available data (Atanas et al. 2023) can be used to test and refine simulated models constructed from IO-functions.
*A concrete potential solution:* Microfluidics can be used to position nematodes onto a microscope stage. Such a system will include a reservoir of age-synchronized animals, in which animals are introduced to chemogenetics reagents. Tubing and channels will then place ten chemogenetically paralyzed adult hermaphrodites next to each other to fill a rectangular field of view (Mondal et al. 2016). Microfluidic technology (San-Miguel and Lu 2013) and automation (Li et al. 2022) are well-established in *C. elegans* and appear to pose no relevant risk to the reverse engineering project.

### (2) Microscopy
*Need:* The nervous system of *C. elegans* is distributed along its body. About two-thirds of the 302 neuronal cell bodies are in a head ganglion, and the rest are split between a tail ganglion and a ventral nerve cord. Animals are about 1 mm in length, and the maximal thickness of the nervous system is about 50 microns. This volume should be imaged at a sampling rate of at least 10 and preferably 100 Hz, requiring some progress relative to today's typically slower rates (Atanas et al. 2023). The experiments require resolving individual neurons and identifying them across experiments. Fortunately, neuronal cell bodies are large enough that diffraction-limited imaging should be sufficient to resolve the relevant signals, particularly if the relevant indicators can be localized to the soma. However, high resolution introduces field-of-view limitations, where typical microscopes have a field-of-view that is too small to fit the entire nematode. Imaging setups will have to be specifically designed for the goal of automated, large-scale systems identification.
*Potential approaches:* Because diffraction-limited microscopy of fluorescent indicators is expected to be sufficient, there are many possible configurations (Lemon and McDole 2020; Balasubramanian et al. 2023). This includes various light-sheet and spinning disk confocal designs that can scan this volume at the required sampling rate. It also includes the more sophisticated two-photon setups that allow exceptional precision in space, e.g., for stimulation.
*A concrete potential solution:* An inverted SCAPE2.0 microscope (Voleti et al. 2019) with a microfluidic stage would be suitable for the optogenetic and imaging approach. The inverted design provides access to liquid handling and stimuli, and the SCAPE2 can scan the required volume (1 x 1 x 0.05 mm for ten immobilized animals) at about 10 Hz. This volume is sufficient to image ten immobilized or one restricted animal.

### (3) Neuron Identification
*Need:* For the recorded activity to be useful, each recorded neuron has to be reliably assigned to one of the 302 neuronal identities. To do so at scale, automated software is needed to identify each neuron from multichannel optical stacks after recording.
*Potential approaches:* A large number of solutions are possible. High-resolution, potential super-resolution, stacks, e.g., from expansion microscopy (Yu et al. 2020), would allow solving the problem by machine learning from large datasets. NeuroPAL (Yemini et al. 2021), a multicolor transgene, allows for nervous-system-wide neuronal identification using a combination of reporters and colors to generate an invariant color map across individuals radically simplifying the identification process. Alternatively, machine learning may enable identification from large sets of multimodal data (Kirillov et al. 2023). This field of identification is quickly maturing.

*A concrete potential solution:* Neuronal identification could build on the tried and tested NeuroPAL (Yemini et al. 2021) strain of transgenic animals and future derivatives using conceptually similar approaches and AI for cell identification. Techniques to automate using NeuroPAL are quickly becoming standard but will still require additional software development (Skuhersky et al. 2022) and the development of novel methods for aligning neurons across thousands of animals. Importantly, it is sufficient to solve the identification problem for most neurons in most animals, and data from unidentified neurons can be discarded.

### (4) Stimulation and Recording

*Need:* The core of reverse engineering a nervous system is figuring out how interactions among components (here neurons) shape neural dynamics and behavior; we need to know how a neuron's output is caused by all its input cells (including those acting through neuromodulators) and how a neuron's output affects other neurons and muscle cells. In other words, we need the interactome, which is a generalization of the connectome. This interactome is complicated because there are both synaptic interactions and neuromodulatory interactions (Ripoll-Sánchez et al. 2022; Beets et al. 2022; Bentley et al. 2016), but there have been recent attempts at collecting a larger number of interactions (Randi et al. 2022a; Uzel, Kato, and Zimmer 2022). These approaches have, however, only revealed the average linear neuron-to-neuron interactions instead of the full nonlinear interactions. We need to know how stimulating (or inactivating) combinations of neurons affect the activity of each neuron. If we had this information, we would be able to calculate each neuron's output given its inputs.

*Potential approaches:* There are many ways of controlling nematode neurons. We can activate them through direct physical effects (Suzuki et al. 2003; O'Hagan, Chalfie, and Goodman 2005; Ramot, MacInnis, and Goodman 2008; Miriam B. Goodman and Sengupta 2019), stimulate them electrically (M. B. Goodman et al. 1998), or stimulate them through optogenetic means (Husson, Gottschalk, and Leifer 2013; Nagel et al. 2005). We can use single-photon or multi-photon optogenetic approaches, trading off price and precision. All of these approaches are well established, but optogenetic methods are particularly scalable. There are also many ways of recording from nematode neurons. We can record from them electrically (M. B. Goodman et al. 1998), with calcium (Kerr et al. 2000), or by voltage imaging with genetically encoded sensors (Azimi Hashemi et al. 2019). Neurons are the most obvious target for stimulation, but a complete interactome might require stimulating other cells, such as glia (Perea et al. 2014). There are a large number of interactions to measure. The main question is which kinds of approaches can be applied to very large datasets derived from large numbers of individuals.

*A concrete potential solution:* To simplify the optogenetic access and experimental design, transgenic nematodes could be generated that stochastically express an optogenetic activator in a random subset of neurons (Aoki et al. 2018; Pospisil, Aragon, and Pillow 2023). Given the experience with Brainbow, this is doable and presents minimal risk. For example, if each neuron has a 1/150 chance of expressing the activator, we expect most (~80%) animals to express the activator in one, two, or three neurons. A few animals express it in more than three neurons, while 10-15% of animals will not express the activator in any neuron and will serve as an internal control group. Providing a single-photon optogenetic stimulation on only one channel makes experiments far simpler and inexpensive than 2-photon targeted stimulation. Light activation should be as brief as possible and as low in intensity as possible to avoid polysynaptic recruitment and overinterpretation (e.g., Penfield and Boldrey, 1937). Another approach to reducing intrinsic activity and polysynaptic activation is to subtly inactivate all neurons and muscle cells by making use of transgenic animals expressing a histamine-gated chloride channel (Pokala et al. 2014). Genetically encoded calcium indicators such as GCaMP (Inoue 2021) are a reasonable proxy for neuronal activity. Voltage imaging of entire brains and nervous systems is now becoming feasible, thanks to new indicators and microscopes (Wang et al. 2023). The specific calcium indicator should be selected for spectral separation from NeuroPAL and the optogenetic illumination.

### (5) Annotated Connectomes

*Need:* To further understand the interactome data and possibly predict neuronal function, it will be beneficial to have a specific molecularly annotated connectome (Bhattacharya et al. 2019; Taylor et al. 2021); in other words, to know the whole structure of the nervous system of animals from which we collected input-output data, but also where key molecules are expressed. This kind of information can significantly cut down on the set of models we need to consider.

*Potential approaches:* Serial electron microscopy and several super-resolution approaches can be adapted to image the morphology of the nervous system of collected and fixed animals. These are mostly low-throughput methods. Expansion microscopy is the only approach we are aware of that could provide molecular annotation in addition to morphological data and at a relatively high throughput (Alon et al. 2021; Shen et al. 2020; Tavakoli et al. 2024).

*A concrete solution:* After stimulation and recording, a fraction of the animals could be extruded from the microfluidic device, and expanded (Yu et al. 2020) to identify a host of molecules in 3 dimensions, along with the complete connectome, which, in turn, will require advances in machine learning and computer vision for automatic reconstruction and proofreading.

### (6) Relevant behavior

*Need:* A good simulation of the nervous system should also be able to do what evolution shaped it to do: generate the rich set of behaviors of *C. elegans* (Hart 2006). A necessary prerequisite, then, is to capture and quantify behavior at a resolution that preserves the postural, spatial, and temporal dimensions that the nervous system acts upon, as well as the environmental context in which that happens. A key consideration should thus be a choice of a range of behaviors with relatively simple world interactions, ideally a set of behaviors that have relatively shared inputs and outputs. The goal is to obtain a holistic view of behavior that includes the environment, sensors, neural dynamics, effectors, and their interactions.

*Potential approaches:* There is a need to start relatively simple but a real success should be able to deal with a broad range of behaviors. As such, starting with spontaneous and motivated behaviors that work with individual worms appears to be the best place, even though many of the relevant behaviors for *C. elegans* have a social dimension that one should ultimately understand. One would like to densely sample relevant behavioral states (e.g., roaming and dwelling) while broadly sampling the full repertoire of behavioral states (Hart 2006; Datta et al.

2019). Quantification of these behaviors can be facilitated by existing tracking and analysis software (Husson et al. 2013; Barlow et al. 2022; Hebert et al. 2021) for the centerline, but finer-scale behaviors might require full body shape with body-part segmentation (Deserno and Bozek 2023), possibly in 3D (Ilett et al. 2023).

*A concrete solution:* Initially focus on interpretable single animal behaviors, such as locomotion and chemotaxis. These behaviors also share important inputs, outputs, and behavioral states. Later, one can move to more open-ended behaviors, including multiple animals and three-dimensional locomotion. Existing tracker software (Javer et al. 2018) and hardware can be adapted to track single or multiple animals in microfluidic devices or freely moving. All behavioral data should be stored in the community standard format, WCON (Javer et al. 2018).

### (7) Quality control

*Need:* Countless variables affect the results of the experiments we sketch here. These range from the genetic background of the nematodes (Bargmann 1998; Hobert 2013) to the properties of the microscopes (Marblestone et al. 2013) and the expression patterns of the indicators (Schmitt et al. 2012). As such, careful monitoring of the overall pipeline is essential. The results must be comparable to pool them.

*Potential Approaches:* There are two kinds of well-established approaches. One, as in the case of the International Brain Lab (Wool and International Brain Laboratory 2020) is to establish standards that are held up in many labs working in parallel. The other, as in the Allen Institute, is to establish one central location where large-scale experiments are made with a team that is very much centered around quality control. A centralized approach might introduce site-related idiosyncrasies that are hard to control and disambiguate.

### (8) Model fitting

*Need:* We need to be able to fit models to the data that are sufficient to faithfully simulate inputs, the dynamics of neural activity and outputs, and the resulting behaviors with and without perturbations, and establish that they generalize to other contexts.

*Potential approaches:* Even in a simple nervous system, building a faithful emulation of the brain and behavior will require complex compositions of nonlinear functions. Thus, one approach would entail fitting deep learning models that could capture these nonlinearities (Tanaka et al. 2019; Benjamin et al. 2018; Yamins and DiCarlo 2016; Doerig et al. 2023). In all cases, we would have to fit models where we estimate neuron outputs $y_j(t)$ as a function of neuron inputs $x_i(t)$. Alternatively, embodied NeuroAI approaches could also be used to bridge neural activity function modeling and behavior, including through the use of connectome-constrained topology (Lappalainen et al. 2023; Cowley et al. 2024) and through the use of deep reinforcement learning and biomechanical simulation that construct the problem as that of inferring a control policy driven by a neural controller (Aldarondo et al. 2024; Melis, Siwanowicz, and Dickinson 2024).

*A concrete solution:* Starting with a deep learning system based on transformers (Vaswani et al. 2017), one could learn a set of descriptors for each input neuron's state $x_i(t)$ using a unit-level tokenization approach (Azabou et al. 2023). The overall system can then be simulated by first learning to predict neural activity and movement from sensory inputs fed into the model.

**(9) Open Science**
*Need:* Such a project will require a concentration of effort and resources that can be best supported by an open community. A community may also provide quality assurance - the data may have problems we do not anticipate.
*Potential solutions:* To produce a governance structure that minimizes that risk or to be absolutely open.
*A concrete solution:* Doing both. It is important to engage the relevant community to guide the approaches and open science has been the standard approach in the *C. elegans* field. This effort can be an opportunity to practice and demonstrate radically open science in which code and data are made available within a month of the time they are developed or collected and analyzed.

**(10) Diverse Science**
*Need:* This project will require the integration of scientists with expertise in instrumentation, genetic analysis, molecular tools, behavioral neuroscience, cellular biophysics, computation, data science, and theory.
*Potential solutions:* Intentionally build an inclusive community of scientists and embed a tradition of design and data review into teamwork. Empower team members to cross-train in multiple disciplines.

**Power calculations**
It is notoriously difficult to estimate statistical power in the context of machine learning applications. However, there are two things that we can do. (1) We can produce a simulation that models the relevant problem and see how well we can reverse engineer that simulation to obtain useful intuition. (2) We can use special cases, such as linear models with nonlinear basis functions, to obtain closed-form estimates: (1) One simulation strategy is to produce virtual connectomes of varying resolution (different numbers of neurons and different densities of connections between them) and varying complexity. For example, to change the number of input parameters to each neuron and how much activity history a neuron takes as input — whereas a perfectly Markovian neuron takes only the latest state of the network as input. Of course, real neurons fall in the middle of this range, though we do not know where. It is possible to empirically evaluate how much recording time is required (Fig. **2**). (2) We find with considerable amounts of uncertainty (from the complexity of *C. elegans* neurons) that we will need to continuously record from the entire *C. elegans* nervous systems for the equivalent of about 900,000 seconds or 250 days (see Appendix 1: analytical power calculations), e.g., 1000-second long experiments in 900 animals. As such, the proposed experiments are doable, even if we use conservative calculations.

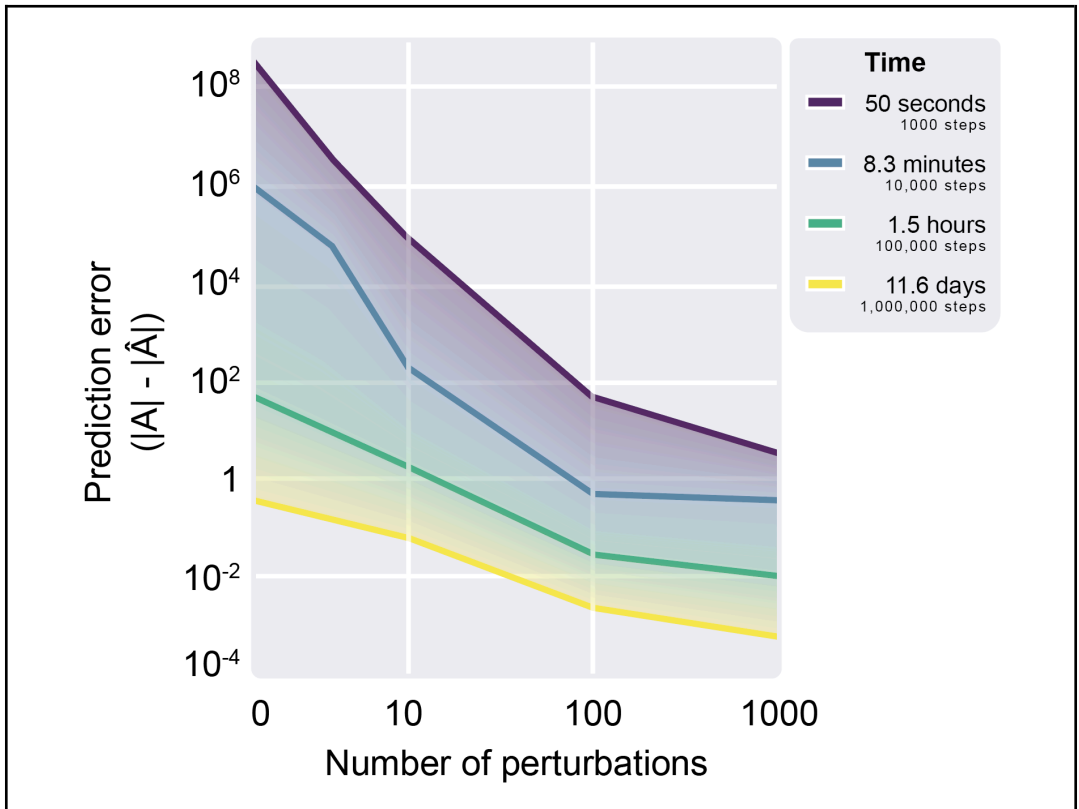

Fig. 2. **A simple linear model of connectome dynamics can be reverse-engineered in the manner described here.** Predictions on a 100-neuron model benefit somewhat from longer-duration recording time but are *dramatically* improved by inducing artificial perturbations during recording. The prediction error is most effectively reduced through a combination of the two. The dynamics of this system are described in *Appendix 2*.

**Beyond *C. elegans***

Ultimately, the value of reverse engineering a nervous system is that it will make reverse engineering nervous systems accessible for comparative approaches. How do neural dynamics change during disease in the nervous system? How can we scale this approach to larger and more nervous systems? How should the methods be adapted and how do the resulting models differ across species? Just like the sequencing of the first genome enabled the Human Genome Project and the sequencing of the genomes of many humans, reverse engineering a nervous system promises to open the floodgates to reverse engineering many other nervous systems (Vogelstein et al. 2016).

**Generalizing to larger nervous systems**

Reverse engineering the *C. elegans* nervous system should just be a step in the path toward understanding nervous systems that are more like our own. There are of course many differences, in terms of size (human brains with about 86 billion neurons vs 302 in *C. elegans*), in terms of computational primitives (most cells in *C. elegans* do not spike at the millisecond timescale; (M. B. Goodman et al. 1998; Jiang et al. 2022; Q. Liu et al. 2018; Lockery and

Goodman 2009), and in terms of complexity of brain organization. However, we can sketch what a path toward larger brain understanding can look like. Power calculations make the identification of the IO-function of cortical neurons in humans that have an order of 10,000 inputs from stimulation seem infeasible. However, through the work that we outline here, we will not merely establish causal interactions between *C. elegans* cells, we will also establish how to predict them from annotated connectomes. Annotated connectomes can readily be scaled to much bigger nervous systems (Glaser et al. 2024). Moreover, live imaging of large volumes at high speeds may also be scalable, with novel innovations such as chemical modification of the transparency of living brain tissue (Talei Franzesi et al. 2024; Inagaki et al. 2024). We also expect that there are some principles (some known, most probably still unknown) that may carry over from small nervous systems to larger ones. The important thing is that the work on reverse engineering *C. elegans* will provide a Rosetta-stone-like translation between the language of annotated connectomes and functional properties. This insight into the causal interactions and this ground truth-based approach can then be generalized. For example, we could analyze the kinds of interactions in human brain organoids producing ground truth causality and then scale up from there. But we will need the whole nervous system scaling to know how to put together such local information with other factors to get at the causal flow of information in larger nervous systems.

**Conclusions**
Here we have outlined how reverse engineering the entire sensory-neuro-motor system of *C. elegans* may be possible, producing, for the first time, an understanding of how an entire nervous system computes and drives behavior. Reverse engineering the causal interactions in the nervous system of *C. elegans* promises to establish in silico simulations as a way of accelerating neuroscience. It promises to teach us how to run experiments that reveal causal mechanisms of neural circuits. We may even learn how to build powerful, error-robust, energetically-economical, synaptically-compact AI systems. A concerted, open effort would galvanize the development of approaches to help understand larger nervous systems, like our own, promising a new era of causal neuroscience. The time has come to reverse engineer the *C. elegans* nervous system.

**Appendix 1: analytical power calculations**

*General setting.* We need to describe the dependency of the neuron's output on its inputs which we can formalize as: $y = f(x)$. Let there be $L$ noisy observations, abstracting away time, of the form $y_i = f(x_i) + \eta_i$ with isotropic Gaussian noise $\eta_i \sim N(0, \sigma^2 I)$ as neural observations generally have channel noise. In this context, $I$ represents the identity matrix and $\sigma^2$ denotes the noise variance. In this setting, we want to ask how well we can approximate $f(x)$ by function fitting on limited amounts of data. We want to describe this function so that we can predict it for all possible inputs $x$. This thus covers all possible behaviors of the nervous system for any stimuli and behaviors as well as the response to any internal perturbations. This setting protects us from having a fragile model that only performs well inside the specific contexts studied in an experiment.

**Analytical power calculations**

*Linear setting*
Instead of the general setting, let us operate in one in which we can describe the function as a sum of $K$ terms of relevant basis functions:

$$f(x) = \sum_{k=1}^{K} W_k g_k(x)$$

With a set of appropriate basis functions $g_k(x)$, which may, e.g. describe synapse-synapse interactions or local dendrite interactions, and weights $W_k$ which describe how important each of the basis functions is. Now, to be clear, real neuron transfer functions can not meaningfully be written in this form, for example, because they have an output nonlinearity. However, while output nonlinearities can e.g., set half of the outputs to be zero, they are unlikely to massively affect the power calculations because real neural output nonlinearities tend to be relatively simple smooth functions (Kato et al. 2014). Importantly, in this scenario, due to the linear nature of the identification problem, we can use the well-established theory for linear systems identification to obtain solid intuitions.

How many such basis functions should we need for a neuron? No one knows. In a super simplistic world, neurons could mostly be linear, in which case we would just have one basis function for each synapse, and we could get away with *K=30*. In a super complicated world, every synapse could be multiplied together into a basis function, and we would need *K>>1 million*. However, in reality, complexity will be way higher than linear as we know that the neuromodulators have major modulating effects. But clearly, we do not expect all combinations to meaningfully interact, e.g. because many synapses are far away from one another. As such, we may believe that the right *K* will be somewhere between *K=1,000*, allowing interactions between any pair of synapses, and *K=100,000* allowing many three-way interactions. We will thus use *K=10,000* as our estimate, knowing full well that uncertainty about this number of parameters needed to describe neurons is high.

Because noise is isotropic in our system, we can use whatever basis function system we like to do our analysis. Importantly, in the basis function of the singular vectors (components) of the $E[xx']$ system (~the system discovered by PCA), the dimensions stop being coupled to one another, and we can view the identification problem as $K$ uncoupled estimates. Instead of estimating the weights in the inconvenient original coordinate system of $X$ where all dimensions have the same noise but complex covariance structure, we will thus estimate all our weights in the convenient coordinate system of the relevant components, where all dimensions are uncorrelated but have different noise levels. We will now call the transformed input activities with means subtracted $x_i$ and whenever we use indices $i$, we imply that we are in this transformed coordinate system. Importantly they are now whitened, and have unit variance and zero covariance with one another.

In that coordinate system, we have our weight estimates:

$$\hat{\beta}_i = <x_i y> / <x_i x_i>$$

Now, in this estimate, errors come from misestimations of the first and the second term. However, when whitening the signals, the $x_i$ are being divided by $PC_i$. And for the difficult-to-estimate dimensions with $k >> 1$, our error estimate is entirely dominated by the second term in the standard error expansion equation. This term has a noise level of

$$\sigma_i \approx \frac{\sigma}{PC_i \sqrt{L}}$$

If all PCs are of the same size (say 1), then we obtain the well-known result:

$$||\beta - \hat{\beta}||_2 = \sigma \frac{\sqrt{K}}{\sqrt{L}}$$

But if the PCs are distinct as they always are in practice, we instead obtain a variance correction.

In practice, PC spectra are extremely heavy-tailed. If we record even just a few hundred neurons, we generally find that the smallest PCs are smaller than the largest PCs by a factor of thousands. As such, estimation becomes very difficult, in particular given that $\sigma \approx PC_1$ in most systems. The intuition for all this is simple, small singular values have only a tiny bit of associated variance. For example, if we have two convergent input neurons that are strongly correlated with one another then the weight associated with the first PC (the average of the two neurons) is easy to estimate while the weight associated with the second PC (the difference of their contributions to their shared target) is hard to estimate as it hard to assign credit to either of the neurons.

This is where stimulation comes in. If we randomly stimulate in the space of the $g_i$ then we are effectively adding the identity matrix to the covariance matrix. Stimulation thus makes all singular values become similar to one another and reduces the difference between the large and the small singular values. In relevant simulations, stimulation can typically get this correction factor (condition number $c$) to be roughly 1 (or at least not larger than 3); see below.

Let us go back to the example of the two correlated input neurons. The stimulation will make them less correlated, making it easier to assign credit to either of the inputs.

So the calculations of power are relatively simple from a mathematical perspective. Suppose an analysis of one neuron with just one input will take, say, 10 seconds (enough for order 100 observations, which we will call $\Delta t_0$) to identify the transfer function. If a neuron has K parameters, we will instead need $c^2 K \Delta t_o$. We assume a worst-case *c*=3 and hence, we would need 900,000 seconds = 15,000 minutes ≈ 250 days. In other words, we would need massive but tractable amounts of data.

**Appendix 2: empirical power calculations**

Here we compute empirical results of a timeseries linear dynamical model on a large sample (N≈1000) of randomly generated connectomes. Dynamics at time-step *t+1* are defined as:

$$Y_t = logistic((A \times X_t + \sigma_t) - \mu)$$

Where A is the (static) connectome adjacency matrix, $X_t$ is the input-state at time t, σ is a variability (standard deviation) which is generated anew at each timestep, and μ a threshold/offset value that remains constant during the simulation.

Taking from biology the understanding that such dynamics operate at the millisecond timescale, we can calculate that one timestep in our simulation corresponds roughly to 20 ms of real-world wallclock time, and therefore a simulation of 1,000 timesteps corresponds to 20 seconds (the purple line in **Figure 2**).

We randomly sample the number of perturbations and compute the recording time to evenly space these perturbations across the duration. Connectivity (density of A) is sampled as a uniform value between .01 and 0.5, broadly accommodating the synaptic density of *C. elegans* (0.06) (Matelsky et al. 2021). The σ parameter is randomly sampled along the log distribution from [0.01-10.0). μ was set to 0.5 for these simulations, and the scale of the perturbations, as a ratio of the number of neurons randomly influenced in each simulation (one value per simulation) was randomly uniformly sampled on the interval [0.01-0.5).